\documentstyle[12pt]{article}
\begin{document}
\parskip=5pt plus 1pt minus 1pt

\begin{flushright}
{\bf Preprint LMU-99-13} \\
{\small September 1999}
\end{flushright}

\vspace{0.2cm}

\begin{center}
{\Large\bf $CP$ Violation and the Light Quark Sector}
\footnote{Invited talk given at the 13th Rencontres de Physique
de la Vallee d'Aoste, {\it Results and Perspectives in Particle
Physics}, La Thuile, Aosta Valley, Italy, March 1999}
\end{center}

\vspace{0.5cm}

\begin{center}
{\bf Harald Fritzsch}  \footnote{Electronic address:
bm@hep.physik.uni-muenchen.de} \\
{\it Sektion Physik, Universit$\ddot{a}$t M$\ddot{u}$nchen,
Theresienstrasse 37A, 
80333 M$\ddot{u}$nchen, Germany}
\end{center}

\vspace{2cm}

\begin{abstract}
In view of the observed strong hierarchy of quark 
masses, we discuss a new description of flavor mixing
which is particularly suited for models of quark mass
matrices based on flavor symmetries. The necessary and
sufficient conditions for $CP$ violation are clarified.
The emergence of $CP$ violation 
is primarily linked to a large phase difference (near $90^{\circ}$)
in the light quark sector. 
The unitarity triangle is determined by the mass ratios 
of the light quarks. We conclude that the unitarity 
triangle should be close or identical to a rectangular triangle,
and $CP$ violation is maximal in this sense. 
\end{abstract}

\newpage

\section{Introduction}
\setcounter{equation}{0}

A deeper understanding of flavor mixing and $CP$
violation, observed in the weak interactions,
remains one of the major challenges in particle physics. In the
standard electroweak theory with three quark families
the phenomenon of flavor mixing is described by a $3\times 3$
unitary matrix, which can be expressed in terms of four 
independent parameters,
usually taken to be three rotation angles and one complex phase.
There seems no way to obtain any further
information about these parameters within the standard model. 
Any attempt to do so would require new physical inputs 
which are beyond the standard model.

At the present time it seems hopeless to find a complete
solution to the fermion mass and flavor mixing problem by
theoretical insight alone. One can hope, however, to detect a
specific order in the tower of fermion masses and the four 
parameters of quark flavor mixing, especially in observing
links between the parameters of the flavor mixing and the mass
eigenvalues. That such links should exist, seems obvious to us.
Like in any quantum mechanical system the mixing pattern of the
states will influence the pattern of the mass eigenvalues, and
vice versa. One possible way to make these links more
transparent is to look for specific symmetry limits, e.g.,
by setting parameters, which are observed to be small, to
zero and to study the situation in the symmetry limit first.
Following such an approach, we shall demonstrate that (a) 
a specific description of quark flavor mixing can be derived,
(b) two of the three flavor mixing angles are related 
directly to the quark mass ratios $m_u/m_c$ and $m_d/m_s$,
and (c) the unitarity triangle of quark mixing
related to $CP$ violation in
$B$-meson decays is fixed in terms of these mass ratios and the 
modulus of the Cabibbo transition element $|V_{us}|$.
Furthermore we shall give arguments why an inner angle 
of the unitarity triangle (angle $\alpha$)
should be equal to $90^{\circ}$ or close to $90^{\circ}$ \cite{FX99}.

The ``standard'' parametrization of the flavor mixing matrix
(advocated by the Particle Data Group \cite{PDG98}) and
the original Kobayashi-Maskawa parametrization \cite{KM}
were introduced
without taking possible links between the quark masses
and the flavor mixing parameters into account. The 
parametrization introduced some time ago \cite{FX97,FX98} 
is based on such a connection, although the
specific relations between flavor mixing angles and
quark masses might be more complicated than 
commonly envisaged. It is a parametrization which allows 
to interpret the phenomenon of flavor mixing as an 
evolutionary or tumbling process. In the limit in which the masses 
of the light quarks $(u,d)$ and the medially light 
quarks $(c,s)$ are set to zero, while the heavy quarks
$(t,b)$ acquire their masses, there is no flavor mixing
\cite{F87}. Once the masses of the $(c,s)$ quarks are
introduced, while the $(u,d)$ quarks remain massless, the
flavor mixing is reduced to an admixture between two families,
described by one angle $\theta$. As soon as the $u$-
and $d$-quark masses are introduced as small perturbations,
the full flavor mixing matrix involving a complex 
phase parameter and two more mixing angles 
$(\theta_{\rm u}, \theta_{\rm d})$ appears.
These angles can be interpreted as rotations between
the states $(u,c)$ and $(d,s)$, respectively. 
In either the ``standard''
parametrization or the Kobayashi-Maskawa representation,
however, such specific limits are difficult to
consider. For this reason we proceed to describe the flavor
mixing by use of the parametrization given in Ref. \cite{FX97}.

\section{The flavor mixing matrix}
\setcounter{equation}{0}

In the standard electroweak theory or those extensions which have no
flavor-changing right-handed currents, it is always
possible to choose a basis of flavor space in which 
the up- and down-type quark mass matrices are 
hermitian. Without loss of any generality the (1,3) and
(3,1) elements of both mass matrices
can further be arranged, through a common unitary
transformation, to be zero \cite{FX97}. 
Then one is left
with hermitian quark mass matrices of the form
\begin{equation}
M_{\rm q} \; =\; \left (\matrix{
E_{\rm q}	& D_{\rm q}	& {\bf 0} \cr
D^*_{\rm q}	& C_{\rm q}	& B_{\rm q} \cr
{\bf 0}		& B^*_{\rm q}	& A_{\rm q} \cr}
\right ) \; ,
\end{equation}
where q = u (up) or d (down), and the hierarchy
$|A_{\rm q}| \gg |B_{\rm q}|, |C_{\rm q}|
\gg |D_{\rm q}|, |E_{\rm q}|$ is generally expected.
In this basis, there is no direct mixing between 
the heavy $t$ (or $b$) quark and the light $u$ (or $d$)
quark in $M_{\rm u}$ (or $M_{\rm d}$), i.e., the
quark mass matrix is close to the well-known form of
``nearest-neighbour'' interactions \cite{F78}.

A mass matrix of the type (2.1) can in the absence 
of complex phases be diagonalized by a $3\times 3$
orthogonal matrix, described only by two rotation
angles in the hierarchy limit of quark masses \cite{F79}. 
First, the off-diagonal element $B_{\rm q}$ is rotated away 
by a rotation matrix $R_{23}$
between the second and third families.
Then the element $D_{\rm q}$ is rotated away by a
transformation $R_{12}$ between the first and second families. 
No rotation between the first and third families is
necessary in either the limit $m_u\rightarrow 0$,
$m_d\rightarrow 0$ or the limit $m_t\rightarrow
\infty$, $m_b\rightarrow \infty$. Lifting such a
hierarchy limit, which is not far from the reality,
one needs an additional transformation
$R_{31}$ with a tiny rotation angle 
to fully diagonalize $M_{\rm q}$.
Note, however, that the rotation sequence
$(R^{\rm u}_{12} R^{\rm u}_{23}) (R^{\rm d }_{12}
R^{\rm d }_{23})^{\rm T}$ is enough to describe the $3\times 3$ real
flavor mixing matrix, as the effects of $R^{\rm u}_{31}$
and $R^{\rm d}_{31}$ can always be absorbed into
this sequence through redefining the relevant rotation
angles. By introducing a complex phase angle into 
the rotation combination $(R^{\rm u}_{23}) (R^{\rm d}_{23})^{\rm T}$,
we finally arrive at the following representation of
quark flavor mixing \cite{FX97}:
\begin{eqnarray}
V & = & \left ( \matrix{
c_{\rm u}       & s_{\rm u}     & 0 \cr
-s_{\rm u}      & c_{\rm u}     & 0 \cr
0       & 0     & 1 \cr } \right )  \left ( \matrix{
e^{-{\rm i}\varphi}     & 0     & 0 \cr
0       & c     & s \cr
0       & -s    & c \cr } \right )  \left ( \matrix{
c_{\rm d}       & -s_{\rm d}    & 0 \cr
s_{\rm d}       & c_{\rm d}     & 0 \cr
0       & 0     & 1 \cr } \right )  \nonumber \\ \nonumber \\
& = & \left ( \matrix{
s_{\rm u} s_{\rm d} c + c_{\rm u} c_{\rm d} e^{-{\rm i}\varphi} &
s_{\rm u} c_{\rm d} c - c_{\rm u} s_{\rm d} e^{-{\rm i}\varphi} &
s_{\rm u} s \cr
c_{\rm u} s_{\rm d} c - s_{\rm u} c_{\rm d} e^{-{\rm i}\varphi} &
c_{\rm u} c_{\rm d} c + s_{\rm u} s_{\rm d} e^{-{\rm i}\varphi}   &
c_{\rm u} s \cr
- s_{\rm d} s   & - c_{\rm d} s & c \cr } \right ) \; ,
\end{eqnarray}
where $s_{\rm u} \equiv \sin\theta_{\rm u}$,
$c_{\rm u} \equiv \cos\theta_{\rm u}$, etc. The three
mixing angles can all be arranged to lie in the first
quadrant, i.e., all $s_{\rm u}$, $s_{\rm d}$, $s$ and
$c_{\rm u}$, $c_{\rm d}$, $c$ are positive.
The phase $\varphi$ may in general take all values 
between 0 and $2\pi$. Clearly $CP$ violation is present, if 
$\varphi \neq 0$ or $\pi$. 

Although we have derived in a heuristic way the particular
description of the flavor mixing matrix (2.2) from the
hierarchical mass matrix (2.1), we should like to emphasize 
that (2.2) is a possible way to describe any mixing matrix,
one out of nine inequivalent representations classified 
in Ref. \cite{FX98}.

If the phase $\varphi$ in $V$ is disregarded, the resulting rotation matrix 
(obtained from (2.2) for $\varphi =0$) is just the one
used originally by Euler; i.e., the angles $\theta$, $\theta_{\rm u}$
and $\theta_{\rm d}$ correspond to the usual Euler angles \cite{EUL}. Note
that this is not the case for other representations of the flavor
mixing matrix given in the literature \cite{KM,Others}.
The representation given in (2.2) can be interpreted as follows. 
First, a rotation by the angle $\theta_{\rm d}$ takes place in the 
plane defined by the $d$ and $s$ quarks. It is followed by
a rotation (angle $\theta$) in the $b$--$s'$ plane, where $s'$ denotes the
superposition $s' = d \sin\theta_{\rm d} + s \cos\theta_{\rm d}$. At
the same time the orthogonal 
state $d' = d \cos\theta_{\rm d} - s \sin\theta_{\rm d}$
is multiplied by the phase factor $e^{-{\rm i}\varphi}$. Finally a rotation 
(angle $\theta_{\rm u}$) is applied in the 
$1$--$2$ plane (about the new third axis). 

The sequence of rotations corresponds just to the Euler sequence \cite{EUL}:
$R_{12} R_{23} R^{\rm T}_{12}$.
On the other hand, the original Kobayashi-Maskawa 
representation \cite{KM} corresponds to the sequence 
$R_{23} R_{12} R^{\rm T}_{23}$, while the ``standard'' 
representation \cite{PDG98}
corresponds to the sequence $R_{23} R_{31} R_{12}$ (see also
the classifications given in Ref. \cite{FX98}). Although all
descriptions of the flavor mixing matrix are mathematically
equivalent, we emphasize that the Euler 
sequence $R_{12} R_{23} R^{\rm T}_{12}$ is physically of 
particular interest, as it involves the rotation matrices $R_{12}$
and $R^{\rm T}_{12}$, which describe the rotations in the light
quark sector, in a symmetric way.
Since the flavor mixing matrix
acts between the quark mass eigenstates ${\cal U} = (u, c, t)$ and
${\cal D} = (d, s, b)$, one could absorb the two $R_{12}$ rotations
in a redefinition of the quark fields. 
The charged weak transition term can be rewritten as follows:
\begin{equation}
\overline{\cal U}_{\rm L} ~ V ~ {\cal D}_{\rm L} \; =\; 
\overline{(u, ~ c, ~ t)}^{~}_{\rm L} ~ V ~ \left ( \matrix{
d \cr s \cr b \cr} \right )_{\rm L} \; =\;  
\overline{(u', ~ c', ~ t)}^{~}_{\rm L} \left ( \matrix{
e^{-{\rm i}\varphi}   & 0     & 0 \cr
0       & c     & s \cr 
0       & -s    & c \cr} \right ) 
\left ( \matrix{
d' \cr s' \cr b \cr} \right  )_{\rm L} \; ,
\end{equation}
where $u' = u \cos\theta_{\rm u} - c \sin\theta_{\rm u}$ and 
$c' = c \cos\theta_{\rm u} + u \sin\theta_{\rm u}$.
Thus the angles $\theta_{\rm u}$ and $\theta_{\rm d}$ describe the
corresponding rotations in the $(u, c)$ and $(d, s)$ systems.

We should like to emphasize that the angles $\theta_{\rm u}$ and
$\theta_{\rm d}$ can directly be measured from weak decays of
$B$ mesons and from $B^0$-$\bar{B}^0$ mixing.  
An analysis of the present experimental
data yields \cite{PRS98}: $\theta_{\rm u} = 4.87^{\circ} \pm 0.86^{\circ}$
and $\theta_{\rm d} = 11.71^{\circ} \pm 1.09^{\circ}$. 
Due to the symmetric structure of our mixing matrix (2.2),
we are able to interpret the $\theta_{\rm d}$ and $\theta_{\rm u}$
rotations as specific transformations of the corresponding
mass eigenstates. Such an interpretation is 
not possible for the third rotation given by $\theta$, measured to
be $2.30^{\circ} \pm 0.09^{\circ}$ \cite{PRS98}. This rotation takes
place between the third family of the massive quarks and the $c'$
and $s'$ states. 
One interpretation would be to associate the rotation of $\theta$
with a transformation among $b$ and $s'$. Another possibility is to
describe the effect as a rotation among $t$ and $c'$. However, one
could also write $\theta$ as a difference of two other angles, and describe
the mixing effect as a combination of a rotation in the $(b, s')$
system and a rotation in the $(t, c')$ system. Thus a unique
interpretation 
does not exist. We remark that the asymmetry between the $\theta$
rotation on the one hand and the $\theta_{\rm u}$ and $\theta_{\rm d}$
rotations on the other hand is a direct consequence of our flavor
mixing matrix (which is in turn related to the hierarchical structure
of the mass spectrum) and is primarily linked to the fact that there exist 
three different quark families.

As summarized in Refs. \cite{FX97,FX98}, the
new parametrization (2.2) has a number of advantages over all
the others in the
study of heavy flavor decays and quark mass matrices.
Its usefulness will be seen more clearly in the present work.
As an example we explore the interesting connection between 
our parametrization (2.2) and the unitarity triangle of quark
mixing defined by the orthogonality relation
\begin{equation}
V^*_{ub}V_{ud} ~ + ~ V^*_{cb}V_{cd} ~ + ~ V^*_{tb}V_{td} \; =\; 0 
\end{equation}
in the complex plane. The inner angles of this triangle,
usually denoted as 
\begin{eqnarray}
\alpha & = & \arg \left ( - \frac{V^*_{tb}V_{td}}{V^*_{ub}V_{ud}}
\right ) \; , \nonumber \\
\beta & = & \arg \left (- \frac{V^*_{cb}V_{cd}}{V^*_{tb}V_{td}}
\right ) \; , \nonumber \\
\gamma & = & \arg \left (- \frac{V^*_{ub}V_{ud}}{V^*_{cb}V_{cd}}
\right ) \; ,
\end{eqnarray}
can be determined from some $CP$ asymmetries in 
$B$-meson decays \cite{BB}. 
The parametrization (2.2) takes an instructive 
leading-order form:
\begin{equation}
V \; \approx \; \left (\matrix{
e^{-{\rm i}\alpha}	& s^{~}_{\rm C} e^{{\rm i}\gamma}	& s_{\rm u} s \cr
s^{~}_{\rm C} e^{{\rm i}\beta}	& 1	& s \cr
-s_{\rm d} s	& -s	& 1 \cr} \right ) \; ,
\end{equation}
where $s^{~}_{\rm C} \equiv \sin \theta_{\rm C} \approx
|s_{\rm u} - s_{\rm d} e^{-{\rm i}\varphi}|$ with $\theta_{\rm C}$
denoting the Cabibbo rotation angle \cite{Cabibbo}. 
Clearly $\alpha \approx \varphi$ holds as a straightforward result of
(2.6). In this approximation $|V^*_{ub}V_{ud}|$, $|V^*_{cb}V_{cd}|$
and $|V^*_{tb}V_{td}|$, the three sides of the unitarity triangle (2.4),
are rescaled to $s_{\rm u}$, $s_{\rm d}$ and $s^{~}_{\rm C}$
respectively. The latter are three sides of a new triangle with
smaller area ($\approx s_{\rm u}s_{\rm d} \sin\alpha/2$), 
which will subsequently be referred to as the ``light-quark triangle''
in the heavy quark limit ($m_t\rightarrow \infty$,
$m_b\rightarrow \infty$).
The values of $\alpha$, $\beta$ and $\gamma$ can
therefore be given in terms of $s_{\rm u}$, $s_{\rm d}$ and 
$s^{~}_{\rm C}$ with the help of the cosine theorem. In
particular, relations like \cite{FX97}
\begin{equation}
\sin\alpha ~ : ~ \sin\beta ~ : ~ \sin\gamma \; \approx \;
s^{~}_{\rm C} ~ : ~ s_{\rm u} ~ : ~ s_{\rm d} \; 
\end{equation}
may directly be confronted with the upcoming data on $CP$
asymmetries in $B$ decays \cite{CDF}. Motivated by these
interesting results, we shall investigate the role that the
light quark sector plays in $CP$ violation for a variety of
realistic textures of quark mass matrices.

\section{Symmetry limits}
\setcounter{equation}{0}

We remark two useful limits of quark masses
and analyze their corresponding consequences on flavor mixing.
In the limit $m_u \rightarrow 0$, $m_d \rightarrow 0$ (``chiral
limit''), where both the up and down quark mass 
matrices have zeros in 
the positions $(1,1)$, $(1,2)$, $(2,1)$, $(1,3)$ and $(3,1)$ (see also 
Ref. \cite{F87}), the flavor mixing angles 
$\theta_{\rm u}$ and $\theta_{\rm d}$ vanish. Only the $\theta$ rotation
affecting the heavy quark sector remains, i.e., 
the flavor mixing matrix effectively takes the form
\begin{equation}
\hat{V} \; =\; \left ( \matrix{
\cos\hat{\theta} & \sin\hat{\theta} \cr
- \sin\hat{\theta} & \cos\hat{\theta} \cr } \right ) \; ,
\end{equation}
where $\hat{\theta}$ denotes the value of $\theta$ which one
obtains in 
the limit $\theta_{\rm u} \rightarrow 0$, $\theta_{\rm d} \rightarrow
0$. We see that $\hat{V}$ is a real orthogonal matrix, arising naturally 
from $V$ in the chiral limit.

The flavor mixing angle $\hat{\theta}$ can be derived from hermitian
quark mass matrices of the following general form (in the limit $m_u
\rightarrow 0$, $m_d \rightarrow 0$):
\begin{equation}
\hat{M}_{\rm q} \; =\; \left ( \matrix{
\hat{C}_{\rm q}	& \hat{B}_{\rm q} \cr
\hat{B}^*_{\rm q} & \hat{A}_{\rm q} \cr } \right ) \; ,
\end{equation}
where $|\hat{A}_{\rm q}| \gg |\hat{B}_{\rm q}| , |\hat{C}_{\rm q}|$; 
and q = u (up) or d 
(down). Note that the phase difference between $\hat{B}_{\rm u}$ and
$\hat{B}_{\rm d}$, denoted as $\kappa \equiv \arg (\hat{B}_{\rm u})
- \arg (\hat{B}_{\rm d})$, has no effect on $CP$ symmetry 
in the chiral limit, but it may affect the magnitude of
$\hat{\theta}$. It is known that current data on the top-quark mass
and the $B$-meson lifetime disfavor the special case $\hat{C}_{\rm u}
= \hat{C}_{\rm d} =0$ for $\hat{M}_{\rm u}$ and 
$\hat{M}_{\rm d}$ (see, e.g., Ref. \cite{FX95}),
hence we take $\hat{C}_{\rm q} \neq 0$ and define a ratio $\hat{r}_{\rm q}
\equiv |\hat{B}_{\rm q}|/\hat{C}_{\rm q}$. We 
can obtain the flavor mixing angle $\hat{\theta}$, in terms of the
quark mass ratios $m_c/m_t$, $m_s/m_b$ and the parameters
$\hat{r}_{\rm u}$, $\hat{r}_{\rm d}$, by diagonalizing the mass
matrices in (3.2). In the next-to-leading order approximation,
$\sin\hat{\theta}$ reads
\begin{equation}
\sin\hat{\theta} \; = \; \left | \hat{r}_{\rm d} \frac{m_s}{m_b} \left (1 -
\hat{\delta}_{\rm d} \right ) ~ - ~ \hat{r}_{\rm u} \frac{m_c}{m_t} \left (1 -
\hat{\delta}_{\rm u} \right ) e^{{\rm i} \kappa} \right | \; ,
\end{equation}
where two correction terms are given by
\begin{eqnarray}
\hat{\delta}_{\rm u} & = & \left (1 + \hat{r}^2_{\rm u} 
\right ) \frac{m_c}{m_t} \; , \nonumber \\
\hat{\delta}_{\rm d} & = & \left (1 + \hat{r}^2_{\rm d} \right ) 
\frac{m_s}{m_b}
\; .
\end{eqnarray}
In view of the fact $m_s/m_b \sim O(10) ~  m_c/m_t$ from current data
\cite{PDG98,Leut}, we find that the
flavor mixing angle $\hat{\theta}$ is primarily linked to $m_s/m_b$
provided $|\hat{r}_{\rm u}| \approx |\hat{r}_{\rm d}|$. Note that in specific 
models, e.g., those describing the mixing between the second and third
families as an effect related to the breaking of an underlying ``democratic
symmetry'' \cite{Democracy,New}, 
the ratios $\hat{r}_{\rm u}$ and $\hat{r}_{\rm d}$ 
are purely algebraic numbers (such as
$|\hat{r}_{\rm u}| = |\hat{r}_{\rm d}| = 1/\sqrt{2}$ 
or $\sqrt{2}$).

For illustration, we take $\hat{r}_{\rm u} = \hat{r}_{\rm d} \equiv
\hat{r}$ to fit the experimental result $\sin\hat{\theta} = 0.040 \pm 0.002$
with the typical inputs $m_b/m_s = 26 - 36$ and $m_t/m_c \sim 250$. 
It is found that the favored value of $|\hat{r}|$ varies in the range
1.0 -- 2.5, dependent weakly on the phase parameter $\kappa$. 

Note that both $m_s/m_b$ and $m_c/m_t$ evolve with the
energy scale (e.g., from the weak scale $\mu \sim 10^2$ GeV
to a superhigh scale $\mu \sim 10^{16}$ GeV,
or vice versa), therefore $\tilde{\theta}$ is a
scale-dependent quantity. 

The limit $m_t \rightarrow \infty$, $m_b \rightarrow \infty$ is
subsequently referred to as the ``heavy quark limit''. In this limit,
in which the $(3,3)$ elements of the up and down  mass matrices formally
approach infinity but
all other matrix elements are fixed, the angle $\theta$ vanishes.
The flavor mixing matrix,
which is nontrivial only in the light quark sector, takes the form:
\begin{eqnarray}
\tilde{V} & = & \left ( \matrix{
\tilde{c}_{\rm u}       & \tilde{s}_{\rm u}     \cr
-\tilde{s}_{\rm u}      & \tilde{c}_{\rm u}     \cr } \right )  
\left ( \matrix{
e^{-{\rm i}\tilde{\varphi}}     & 0     \cr
0       & 1 \cr } \right )  \left ( \matrix{
\tilde{c}_{\rm d}       & -\tilde{s}_{\rm d}    \cr
\tilde{s}_{\rm d}       & \tilde{c}_{\rm d}     \cr } \right )  
\nonumber \\ \nonumber \\
& = & \left ( \matrix{
\tilde{s}_{\rm u} \tilde{s}_{\rm d} + \tilde{c}_{\rm u} \tilde{c}_{\rm d} 
e^{-{\rm i}\tilde{\varphi}} &
\tilde{s}_{\rm u} \tilde{c}_{\rm d} - \tilde{c}_{\rm u} \tilde{s}_{\rm d} 
e^{-{\rm i}\tilde{\varphi}} \cr
\tilde{c}_{\rm u} \tilde{s}_{\rm d} - \tilde{s}_{\rm u} \tilde{c}_{\rm d} 
e^{-{\rm i}\tilde{\varphi}} &
\tilde{c}_{\rm u} \tilde{c}_{\rm d} + \tilde{s}_{\rm u} \tilde{s}_{\rm
d} e^{-{\rm i}\tilde{\varphi}} \cr } \right ) \; .
\end{eqnarray}
where $\tilde{s}_{\rm u} = {\rm sin} \tilde{\theta }_{\rm u},
\tilde{c}_{\rm u} = {\rm cos} \tilde{\theta}_{\rm u}$, etc. The angles
$\tilde{\theta}_{\rm u}$ and $\tilde{\theta}_{\rm d}$ are the 
values for $\theta_{\rm u}$ and $\theta_{\rm d}$
obtained in the heavy quark limit. Since the $(t, b)$ system is decoupled from
the $(c, s)$ and $(u, d)$ systems, the flavor mixing can be described as in
the case of two families. Therefore the mixing matrix $\tilde{V}$ is
effectively given in terms of only a single rotation angle, 
the Cabbibo angle $\theta_{\rm C}$:
\begin{equation}
\sin \theta_{\rm C} = \mid \tilde{s}_{\rm u} \tilde{c}_{\rm d}
~ - ~ \tilde{c}_{\rm u} \tilde{s}_{\rm d} ~ e^{-{\rm i}
\tilde{\varphi}} \mid \; .
\end{equation}
Of course $\tilde{V}(\theta_{\rm C})$ 
is essentially a real matrix, because its complex
phases can always be rotated away by redefining the quark fields.

We should like to stress that the heavy quark limit, which carries
the flavor mixing matrix $V$ to its simplified form $\tilde{V}$, is not far
from the reality, since $1 - c \approx 0.1 \%$ holds \cite{FX97}. 
Therefore $\theta_{\rm u}$, 
$\theta_{\rm d}$ and $\varphi $ are expected to
approach $\tilde{\theta}_{\rm u}$, $\tilde{\theta}_{\rm d}$ and 
$\tilde{\varphi}$ rapidly,
as $\theta \rightarrow 0$, corresponding to
$m_t \rightarrow \infty $ and $m_b \rightarrow \infty$. However, the concrete
limiting behavior depends on the specific algebraic 
structure of the up and down mass matrices.
If two hermitian mass matrices have the parallel hierarchy with texture zeros
in the (1,1) (2,2), (1,3) and (3,1) elements, for example, the magnitude of
$\theta $ is suppressed by the terms proportional to $m_t^{-1/2}$ and
$m_b^{-1/2}$ \cite{F79}; and 
if the (2,2) elements are kept nonvanishing and comparable in magnitude 
with the (2,3) and (3,2) elements, then
$\theta $ is dependent 
on $m_t^{-1}$ and $m_b^{-1}$ \cite{Democracy,New}.

The angles $\tilde{\theta}_{\rm u}$ and 
$\tilde{\theta}_{\rm d}$ as well as the phase
$\tilde{\varphi }$ are well-defined quantities in the heavy quark limit. The
physical meaning of these quantities can be seen more clearly, if we take
into account a specific and realistic model for the Cabibbo-type mixing in
the light quark sector. It is well known that in the absence of the
$u$-quark mass a relation between the Cabibbo angle $\theta_{\rm C}$ 
and the mass
ration $m_d/m_s$ follows, if the quark mass matrices have the structure:
\begin{eqnarray}
\tilde{M}_{\rm u} & = & \left( \matrix{
~ {\bf 0} & ~ {\bf 0} ~ \cr
~ {\bf 0} & ~ m_c \cr } \right) \; , \nonumber \\
\tilde{M}_{\rm d} & = & \left( \matrix{
{\bf 0} & \tilde{B}_{\rm d} \cr
\tilde{B}^*_{\rm d} & \tilde{A}_{\rm d} \cr } \right) \; .
\end{eqnarray}
The diagonalization of $\tilde{M}_{\rm d}$ leads to the relation
${\rm tan } \theta_{\rm C} = \sqrt{m_d/m_s}$ . The texture-zero pattern
of $\tilde{M}_{\rm d}$,
i.e., the vanishing of its (1,1) element, is already 
present in certain classes of
models (see, e.g., Refs. \cite{F77,Weinb}). 
The relation for the Cabibbo angle is known to
agree very well with the experimental observation. For numerical discussions,
we make use of the quark masses $m_u = (5.1 \pm 0.9)$ MeV, $m_d = (9.3 \pm
1.4)$ MeV, $m_s = (175 \pm 25)$ MeV and $m_c = (1.35 \pm 0.05)$ GeV at the
scale $\mu =1$ GeV \cite{Leut}. Then one finds 
$\theta_{\rm C} = 13.0^{\circ } \pm 1.8^{\circ}$ or
$\sin \theta_{\rm C} = 0.225 \pm 0.031$, consistent with the observed
value of $|V_{us}|$ (i.e., $0.217 \leq |V_{us}| \leq 0.224$
\cite{PDG98}).

The situation will change once $m_u$ is introduced, i.e., 
$\tilde{M}_{\rm u}$ takes the same
form as $\tilde{M}_{\rm d}$ given in (3.7). In this case 
the mass matrices result in the
following relation \cite{F79}:
\begin{equation}
\sin \theta_{\rm C} \; = \; \mid R_{\rm u} ~ - ~ R_{\rm d} ~ e^{-{\rm i}
\psi} \mid \; ,
\end{equation}
where
\begin{eqnarray}
R_{\rm u} & = & \sqrt{\frac{m_u}{m_u + m_c}} \, \sqrt{\frac{m_s}{m_d + m_s}}
\;\; , \nonumber \\
R_{\rm d} & = & \sqrt{\frac{m_c}{m_u + m_c}} \, \sqrt{\frac{m_d}{m_d +
m_s}} \; \; ,
\end{eqnarray}
and $\psi \equiv \arg (\tilde{B}_{\rm u}) - \arg (\tilde{B}_{\rm d})$ 
denotes the relative phase between the off-diagonal elements
$\tilde{B}_{\rm u}$ and $\tilde{B}_{\rm d}$ 
(in the limit $m_u \rightarrow 0$ this phase can be absorbed through a
redifinition of the quark fields). We find that the same structure for the
Cabibbo-type mixing matrix has been 
obtained as in the decoupling limit discussed
above. If we set
\begin{eqnarray}
\tan \tilde{\theta}_{\rm u} & = & \sqrt{\frac{m_u}{m_c}} \; , \nonumber \\
\tan \tilde{\theta}_{\rm d} & = & \sqrt{\frac{m_d}{m_s}} \; ,
\end{eqnarray}
and $\tilde{\varphi} = \psi$ for (3.6), then the result in (3.8)
and (3.9) can exactly be reproduced.

\begin{figure}[t]
\begin{picture}(400,160)(-90,210)
\put(80,300){\line(1,0){150}}
\put(80,300.5){\line(1,0){150}}
\put(150,285.5){\makebox(0,0){$\sin\theta_{\rm C}$}}
\put(80,300){\line(1,3){21.5}}
\put(80,300.5){\line(1,3){21.5}}
\put(80,299.5){\line(1,3){21.5}}
\put(71,333){\makebox(0,0){$R_{\rm u}$}}
\put(230,300){\line(-2,1){128}}
\put(230,300.5){\line(-2,1){128}}
\put(178,343.5){\makebox(0,0){$R_{\rm d}$}}
\put(107.5,348.5){\makebox(0,0){$\tilde{\varphi}$}}
\end{picture}
\vspace{-2.6cm}
\caption{The light-quark triangle (LT) in the complex plane.}
\end{figure}
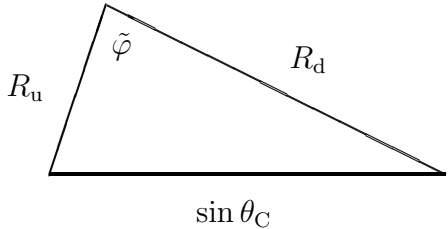
Indeed the relation in (3.6) or (3.8) defines a triangle in the complex plane,
which will be denoted as the ``light-quark 
triangle''(LT). Taking into account the central values of the
Cabibbo angle ($\sin \theta_{\rm C} = |V_{us}| = 0.2205$) and the light
quark mass ratios $\left( m_s / m_d = 18.8 \right. $ and
$\left. m_c /m_u = 265 \right)$, we
can calculate the phase parameter from (3.8) and obtain 
$\tilde{\varphi} = \psi 
\approx 79^{\circ}$. If we allow the mass ratios 
and $\theta_{\rm C}$ to vary in
their ranges given above, then $\tilde{\varphi}$ may vary in the range
$38^{\circ} - 115^{\circ}$. We find that $\tilde{\varphi}$ has a good
chance to be $90^{\circ}$ (see also Ref. \cite{FX95}). 
The case $\tilde{\varphi} \approx 90^{\circ}$ (i.e., the LT is
rectangular) is of special interest, since it
implies that the area of the unitarity triangle of flavor mixing 
takes its maximum value for the fixed quark mass ratios -- in this
sense, the $CP$ symmetry of weak interactions would be maximally violated.

The two symmetry limits discussed above are both not far from
the reality, in which the strong hierarchy of quark masses
($m_u\ll m_c \ll m_t$ and $m_d\ll m_s\ll m_b$) has been
observed. They will serve as a guide in the subsequent 
discussions about generic quark mass matrices and their 
consequences on flavor mixing.

\section{The texture of mass matrices}
\setcounter{equation}{0}

We return to the case of three quark families. 
We adopt a basis of flavor space in which both the up-
and down-type quark mass matrices are hermitian and have
vanishing (1,3) and (3,1) elements, as shown in (2.1).
Such a basis is of special interest in case of a strong mass hierarchy (as
realized by nature), since no explicit mixing between the very massive
$t$ (or $b$) quark and the very light $u$ (or $d$) quark is introduced. 
The mixing can then be regarded as
of the ``nearest neighbour'' type. 
Thus without loss of generality one may discuss the model-independent 
properties of flavor mixing and $CP$ violation on the basis of the mass
matrices (2.1), i.e., 
\begin{eqnarray}
M_{\rm u} & = & \left ( \matrix{
E_{\rm u}       & D_{\rm u}     & {\bf 0} \cr
D^*_{\rm u}     & C_{\rm u}     & B_{\rm u} \cr
{\bf 0}       & B^*_{\rm u}   & A_{\rm u} \cr} \right ) \; ,
\nonumber \\
M_{\rm d} & = & \left ( \matrix{
E_{\rm d}	& D_{\rm d}	& {\bf 0} \cr
D^*_{\rm d}	& C_{\rm d}	& B_{\rm d} \cr
{\bf 0}	& B^*_{\rm d}	& A_{\rm d} \cr} \right ) \; .
\end{eqnarray}
The phases of $D_{\rm u,d}$ and $B_{\rm u,d}$ elements are denoted
as $\phi_{D_{\rm u,d}}$ and $\phi_{B_{\rm u,d}}$, respectively.
The phase differences
\begin{eqnarray}
\phi_1 & = & \phi_{D_{\rm u}} - \phi_{D_{\rm d}} \; , 
\nonumber \\
\phi_2 & = & \phi_{B_{\rm u}} - \phi_{B_{\rm d}} \;
\end{eqnarray}
are the source of
$CP$ violation in weak interactions of quarks.
It is clear that $M_{\rm u}$ and $M_{\rm d}$ consist totally of
twelve parameters.

This general case has been discussed in Ref. \cite{FX99}, where
it is pointed out that the observations indicate that
both $E_{\rm u}$ and $E_{\rm d}$ elements are either very small or zero.
We proceed to specify the general hermitian mass matrices
by taking $E_{\rm q} =0$:
\begin{equation}
M_{\rm q} \; = \; \left ( \matrix{
{\bf 0} & D_{\rm q} & {\bf 0} \cr
D^*_{\rm q} & C_{\rm q} & B_{\rm q} \cr
{\bf 0} & B^*_{\rm q} & A_{\rm q} \cr } \right ) \; .
\end{equation}
In case of two quark families, this is just the
form taken for $\tilde{M}_{\rm d}$ in (3.7). 
As remarked above, the texture zeros
in (1,3) and (3,1) positions can always be arranged. 
Thus the physical constraint is as follows: in the
flavor basis in which (1,3) and (3,1) elements of $M_{\rm u,d}$
vanish, the (1,1) element of $M_{\rm u,d}$ vanishes as well.
This can strictly be true only at a particular energy scale. The vanishing of
the (1,1) element can be viewed as a result of 
an underlying flavor symmetry, which
may either be discrete or continuous. 

In the literature a number of such
possibilities have been discussed (see, e.g., 
Refs. \cite{F79} -- \cite{Zero4}).
Here we shall not discuss further details in this respect, 
but concentrate on the phenomenological
consequences of such a texture pattern.
It is particularly interesting that some predictions of this ansatz
for the mixing angles and the unitarity triangle are approximately
independent of the renormalization-group effects, therefore 
a specification of the energy scale at which the texture of
$M_{\rm u,d}$ holds is unnecessary for our purpose.
We believe that $M_{\rm q}$ given in (4.3) is 
a realistic candidate for the quark mass matrices of a
(yet unknown) fundamental theory responsible for fermion mass generation
and $CP$ violation, and we shall make some further speculations
about this point at the end of this talk.

We define $|B_{\rm q}|/C_{\rm q} \equiv r^{~}_{\rm q}$ for each
quark sector.
The magnitude of $r^{~}_{\rm q}$ 
is expected to be of $O(1)$.
The parameters $A_{\rm q}$, $|B_{\rm q}|$, $C_{\rm q}$ 
and $|D_{\rm q}|$ in (4.3)
can be expressed in terms of the quark mass eigenvalues and $r^{~}_{\rm q}$. 
We obtain the three mixing angles of $V$ as follows:
\begin{eqnarray}
\tan\theta_{\rm u} & = & \sqrt{\frac{m_u}{m_c}} ~ \left ( 1 + \Delta_{\rm u}
\right ) \; , \nonumber \\
\tan\theta_{\rm d} & = & \sqrt{\frac{m_d}{m_s}} ~ \left ( 1 + \Delta_{\rm d}
\right ) \; , \nonumber \\
\sin\theta & = & \left | r_{\rm d} \frac{m_s}{m_b} \left (1 -
\delta_{\rm d} \right ) ~ - ~ r_{\rm u} \frac{m_c}{m_t} \left ( 1 -
\delta_{\rm u} \right ) e^{{\rm i}\phi_2} \right | \; ,
\end{eqnarray}
where the next-to-leading order corrections read
\begin{eqnarray}
\Delta_{\rm u} & = & \sqrt{\frac{m_c m_d}{m_u m_s}} ~ \frac{m_s}{m_b} 
~ \left | {\rm Re} \left [ e^{{\rm i}\phi_1} ~ - ~ \frac{r_{\rm
u}}{r_{\rm d}} \cdot \frac{m_c m_b}{m_t m_s} 
e^{{\rm i}(\phi_1 + \phi_2)} \right
]^{-1} \right | \; , \nonumber \\
\Delta_{\rm d} & = & \sqrt{\frac{m_u m_s}{m_c m_d}} ~ \frac{m_c}{m_t} 
~ \left | {\rm Re} \left [ e^{{\rm i}\phi_1} ~ - ~ \frac{r_{\rm
d}}{r_{\rm u}} \cdot \frac{m_t m_s}{m_c m_b} 
e^{{\rm i}(\phi_1 + \phi_2)} \right
]^{-1} \right | \; ;
\end{eqnarray}
and
\begin{eqnarray}
\delta_{\rm u} & = & \frac{m_u}{m_c} ~ + 
\left ( 1 + r^2_{\rm u} \right ) \frac{m_c}{m_t}
\; , \nonumber \\
\delta_{\rm d} & = & \frac{m_d}{m_s} ~ + \left ( 1 + r^2_{\rm d} 
\right ) \frac{m_s}{m_b}
\; .
\end{eqnarray}
Clearly the result for $\hat{\delta}_{\rm u,d}$ in (3.4) can be
reproduced from $\delta_{\rm u,d}$ in (4.6), if one takes the chiral 
limit $m_u \rightarrow 0$, $m_d \rightarrow 0$. 
We also observe that the phase $\phi_2$ is 
only associated with the small quantity $m_c/m_t$ in $\sin\theta$. 
To get the relationship between $\varphi$ and $\phi_1$ or $\phi_2$,
we first calculate $|V_{us}|$:
\begin{equation}
|V_{us}| \; = \; \left (1 -\frac{1}{2} \frac{m_u}{m_c} - \frac{1}{2}
\frac{m_d}{m_s} \right ) \left | \sqrt{\frac{m_d}{m_s}} ~ - ~
\sqrt{\frac{m_u}{m_c}} ~ e^{{\rm i}\phi_1} \right | \; 
\end{equation}
in the next-to-leading order approximation. Note that this result can
also be achieved from (3.8) and (3.9), which were obtained in the
heavy quark limit. Confronting (4.7) with current data on
$|V_{us}|$ leads to the result $\phi_1 \sim
90^{\circ}$, as we have discussed before. Therefore $\cos\phi_1$ is
expected to be a small quantity.
Then we use (4.20) together with (4.4) and (4.7) 
to calculate $\cos\varphi$.
In the same order approximation, we arrive at
\begin{equation}
\cos\varphi \; =\; \sqrt{\frac{m_u m_s}{m_c m_d}} ~ \Delta_{\rm u} ~ + ~
\sqrt{\frac{m_c m_d}{m_u m_s}} ~ \Delta_{\rm d} ~ + ~ \left (1 -
\Delta_{\rm u} - \Delta_{\rm d} \right ) \cos\phi_1 \; .
\end{equation}
The contribution of $\phi_2$ to $\varphi$ is substantially suppressed at
this level of accuracy. 

For simplicity, we proceed by taking $r_{\rm u} = r_{\rm d} \equiv r$,
which holds in some models with natural flavor
symmetries \cite{Democracy}. Then
$\sin\theta$ becomes proportional to a universal parameter $|r|$.
In view of the fact $m_s/m_b \sim O(10) ~ m_c/m_t$, we find that the result
in (4.5) can be simplified as
\begin{eqnarray}
\Delta_{\rm u} & = & \sqrt{\frac{m_c m_d}{m_u m_s}} ~ \frac{m_s}{m_b} 
~ \cos\phi_1 \; , \nonumber \\
\Delta_{\rm d} & = & 0 \; . 
\end{eqnarray}
Also the relation between $\varphi$ and $\phi_1$ in (4.8) is simplified to
\begin{equation}
\cos\varphi \; =\; \left (1 + \frac{m_s}{m_b} \right ) \cos\phi_1 \; .
\end{equation}
As $m_s/m_b \sim 4\%$, it becomes apparent that 
$\varphi \approx \phi_1$ is a good approximation.
Note that $\phi_1 = \varphi$ holds exactly in the
heavy quark limit, in which $\varphi$ has been denoted
as $\tilde{\varphi}$ (see (3.5) as well as Fig. 1). 
The equality $\phi_1 = \tilde{\varphi}$ follows, 
i.e., both stand for the
phase difference between the mass matrix elements
$D_{\rm u}$ and $D_{\rm d}$. 

Let us calculate the parameter ${\cal J}$ ($= s_{\rm u} c_{\rm u}
s_{\rm d} c_{\rm d} s^2 c \sin\varphi$ \cite{FX97}),
a rephasing-invariant quantity of $CP$ violation 
in the quark sector.
We find that the magnitude of $\cal J$
is dominated by the $\sin\phi_1$ term and receives
one-order smaller corrections from the $\sin (\phi_1 \pm \phi_2)$
terms. As a result, 
\begin{equation}
{\cal J} \; \approx \; |r|^2 \sqrt{\frac{m_u}{m_c}}
\sqrt{\frac{m_d}{m_s}} \left (\frac{m_s}{m_b}\right )^2
\sin\phi_1 \; 
\end{equation}
holds to a good degree of accuracy. Clearly 
${\cal J} \sim O(10^{-5}) \times \sin\phi_1$ with $\sin\phi_1 \sim 1$
is favored by current data.

The result of $\cal J$ in (4.12) might give the impression that
$CP$ violation is absent if either $m_u$ or $m_d$ vanishes. This
is not exactly true, however. If we set $m_u=0$,
$\cal J$ is not zero, but it becomes dependent on $\sin \phi_2$
with a factor which is about two orders of magnitude
smaller (i.e., of order $10^{-7}$):
\begin{equation}
{\cal J} \; \approx \; |r|^2 ~ \frac{m_c}{m_t} \cdot
\frac{m_d}{m_s} \left (\frac{m_s}{m_b} \right )^2
\sin\phi_2 \; .
\end{equation}
Certainly this possibility is already ruled out by experimental data.

Note also that the model predicts 
\begin{eqnarray}
\tan\theta_{\rm u} \; =\; 
\left | \frac{V_{ub}}{V_{cb}} \right | & = &
\sqrt{\frac{m_u}{m_c}} ~ \left (1 + \Delta_{\rm u} \right ) \; ,
\nonumber \\
\tan\theta_{\rm d} \; =\;
\left | \frac{V_{td}}{V_{ts}} \right | & = &
\sqrt{\frac{m_d}{m_s}} ~ \left (1 + \Delta_{\rm d} \right ) \; ,
\end{eqnarray}
a result obtained first by one of us from a
more specific pattern of quark mass matrices \cite{F78}.
In $B$-meson physics, $|V_{ub}/V_{cb}|$ can be determined
from the ratio of the decay rate of $B\rightarrow 
(\pi, \rho) l \nu^{~}_l$ to that of $B\rightarrow D^* l\nu^{~}_l$;
and $|V_{td}/V_{ts}|$ can be extracted from the ratio of
the rate of $B^0_d$-$\bar{B}^0_d$ mixing to that of 
$B^0_s$-$\bar{B}^0_s$ mixing. 

\begin{figure}[t]
\begin{picture}(400,160)(130,210)
\put(300,300){\line(1,0){150}} 
\put(300,300.5){\line(1,0){150}}
\put(370,285.5){\makebox(0,0){$|V_{cd}|$}}
\put(300,300){\line(1,3){21.5}}
\put(300,300.5){\line(1,3){21.5}}
\put(300,299.5){\line(1,3){21.5}}
\put(292,333){\makebox(0,0){$S_{\rm u}$}}
\put(450,300){\line(-2,1){128}}
\put(450,300.5){\line(-2,1){128}}
\put(395,343.5){\makebox(0,0){$S_{\rm d}$}}
\put(315,310){\makebox(0,0){$\gamma$}}
\put(408,309){\makebox(0,0){$\beta$}}
\put(328,350){\makebox(0,0){$\alpha$}}
\end{picture}
\vspace{-2.6cm}
\caption{The rescaled unitarity triangle (UT) in the complex plane.}
\end{figure}
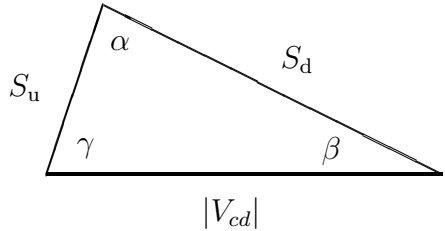

We are now in a position to calculate the unitarity triangle (UT) of
quark flavor mixing defined in (2.8), whose three inner angles are
denoted as $\alpha$, $\beta$ and $\gamma$ in (2.9). Note that three
sides of the unitarity triangle can be rescaled by $V_{cb}^*$ (see
Fig. 2 for illustration). The resultant triangle reads
\begin{equation}
|V_{cd}| \; =\; \left | S_{\rm d} ~ -~ S_{\rm u} ~ e^{-{\rm i}\alpha}
\right | \; ,
\end{equation}
where $S_{\rm u} = |V_{ub}V_{ud}/V_{cb}|$ and $S_{\rm d} =
|V_{tb}V_{td}/V_{cb}|$. After some calculations $S_{\rm u}$, $S_{\rm
d}$ and $\alpha$ are obtained from the above quark mass texture in the
next-to-leading order approximation:
\begin{eqnarray}
S_{\rm u} & = & \sqrt{\frac{m_u}{m_c}} \left ( 1 - \frac{1}{2}
\frac{m_u}{m_c} - \frac{1}{2} \frac{m_d}{m_s} + \sqrt{\frac{m_c
m_d}{m_u m_s}} ~ \frac{m_s}{m_b} ~ \cos\phi_1 + \sqrt{\frac{m_u
m_d}{m_c m_s}} ~ \cos\phi_1 \right ) \; , \nonumber \\ S_{\rm d} & = &
\sqrt{\frac{m_d}{m_s}} \left ( 1 + \frac{1}{2} \frac{m_u}{m_c} -
\frac{1}{2} \frac{m_d}{m_s} \right ) \; ;
\end{eqnarray}
and
\begin{equation}
\sin\alpha \; = \; \left ( 1 - \sqrt{\frac{m_u m_d}{m_c m_s}} ~
\cos\phi_1 \right ) \sin\phi_1 \; .
\end{equation}
A comparison of the rescaled UT in Fig. 2 with the LT in Fig. 1, which
is obtained in the heavy quark limit, is interesting. We find
\begin{eqnarray}
\frac{S_{\rm u} - R_{\rm u}}{R_{\rm u}} & = & \left ( 1 + \frac{m_c
m_s}{m_u m_b} \right ) \sqrt{\frac{m_u m_d}{m_c m_s}} ~
\cos\tilde{\varphi} \; , \nonumber \\ \frac{S_{\rm d} - R_{\rm
d}}{R_{\rm d}} & = & \frac{m_u}{m_c} \; , \nonumber \\
\frac{\sin\alpha - \sin\tilde{\varphi}}{\sin\tilde{\varphi}} & = & -
\sqrt{\frac{m_u m_d}{m_c m_s}} ~ \cos\tilde{\varphi} \; ,
\end{eqnarray}
which are of order $15\% \cos\tilde{\varphi}$, $0.4\%$ and $1.4\%
\cos\tilde{\varphi}$, respectively. Obviously $R_{\rm d} \approx
S_{\rm d}$ is an excellent approximation, and $\alpha \approx
\tilde{\varphi} \approx \varphi$ is a good approximation. As $\varphi$
(or $\tilde{\varphi}$) is expected to be close to $90^{\circ}$,
$R_{\rm u} \approx S_{\rm u}$ should also be accurate enough in the
next-to-leading order estimation.  Therefore the light-quark triangle
is essentially {\it congruent with} the rescaled unitarity triangle!
This result has two straightforward implications: first, $CP$
violation is an effect arising primarily from the light quark sector;
second, the $CP$-violating observables ($\alpha$, $\beta$, $\gamma$)
can be predicted in terms of the light quark masses and the phase
difference between up and down mass matrices \cite{FX95}.  If we use
the value of $|V_{cd}|$, which is expected to equal $|V_{us}|$ within
the $0.1\%$ error bar, then all three angles of the
unitarity triangle can be calculated in terms of $m_u/m_c$, $m_d/m_s$
and $|V_{cd}|$ to a good degree of accuracy.

The three angles of the UT ($\alpha$, $\beta$ and $\gamma$) will be
well determined at the $B$-meson factories,
e.g., from the $CP$
asymmetries in $B_d\rightarrow \pi^+\pi^-$, $B_d\rightarrow J/\psi
K_S$ and $B^{\pm}_u\rightarrow (D^0, \bar{D}^0) + K^{(*)\pm}$
decays \cite{BB}.
The characteristic measurable quantities are $\sin
(2\alpha)$, $\sin (2\beta)$ and $\sin^2\gamma$, respectively. For
the purpose of illustration, we typically take 
$|V_{us}| = |V_{cd}| =0.22$,
$m_u/m_c =0.0056$, $m_d/m_s = 0.045$ and $m_s/m_b = 0.033$ to
calculate these three $CP$-violating parameters from 
the LT and from the rescaled UT separately. 
Both approaches lead to
$\alpha \approx 90^{\circ}$,
$\beta \approx 20^{\circ}$ and
$\gamma \approx 70^{\circ}$,
which are in good agreement with the results obtained from
the standard 
analysis of current data on $|V_{ub}/V_{cb}|$, $\epsilon^{~}_K$,
$B^0_d$-$\bar{B}^0_d$ mixing and $B^0_s$-$\bar{B}^0_s$ mixing
\cite{PRS98}. Note that among three $CP$-violating observables
only $\sin (2\beta)$ is remarkably sensitive to the value of
$m_u/m_c$, which involves quite large uncertainty (e.g., $\sin
(2\beta)$ may change from $0.4$ to $0.8$ if $m_u/m_c$ varies in the
range $0.002 - 0.01$). For this reason 
we emphasize again that the numbers given above can only serve
as an illustration.
A more reliable determination of the 
quark mass values is crucial, in order to test the ans$\rm\ddot{a}$tze 
of quark mass matrices in a numerically decisive way

It is also worth mentioning that
the result $\tan\theta_{\rm d} = \sqrt{m_d/m_s}$ is particularly 
interesting for the mixing rates of 
$B^0_d$-$\bar{B}^0_d$ and $B^0_s$-$\bar{B}^0_s$
systems, measured by $x_{\rm d}$ and $x_{\rm s}$ respectively \cite{PDG98}.
The ratio $x_{\rm s} /x_{\rm d}$ amounts to $|V_{ts}/V_{td}|^2 =
\tan^{-2} \theta_{\rm d}$ multiplied by a factor $\chi_{\rm su(3)} = 1.45
\pm 0.13$, which reflects the $\rm SU(3)_{\rm flavor}$ symmetry breaking
effects. As $x_{\rm d} = 0.723 \pm 0.032$ has been well 
determined \cite{PDG98}, the prediction for the value of $x_{\rm s}$ is
\begin{equation}
x_{\rm s} \; =\; x_{\rm d} ~ \chi_{\rm su(3)} ~ \frac{m_s}{m_d} \; =\;
19.8 \pm 3.5 \; ,
\end{equation}
where $m_s/m_d = 18.9 \pm 0.8$, obtained from the chiral perturbation
theory \cite{Leut}, has been used. This result is certainly consistent
with the present experimental bound on $x_{\rm s}$, i.e.,
$x_{\rm s} > 14.0$ at the $95\%$ confidence level \cite{PDG98}. 
A measurement 
of $x_{\rm s} \sim 20$ may be realized at the forthcoming
HERA-$B$ and LHC-$B$ experiments.

\section{Discussions and conclusion}
\setcounter{equation}{0}

We have studied the phenomena of quark flavor mixing and $CP$
violation in the context of generic hermitian mass matrices.
The necessary and sufficient conditions for $CP$ violation in
the standard model 
have been clarified at both the level of quark mass matrices
and that of the flavor mixing matrix. Our particular observation
is that $CP$ violation is primarily linked to a phase 
difference of about $90^{\circ}$ in the light quark sector,
and this property becomes most apparent in the new 
parametrization (2.2). To be more
specific, we have analyzed a realistic pattern of quark mass
matrices with four texture zeros and given predictions
for the flavor mixing and $CP$-violating parameters. 
The approximate congruency between the light-quark triangle (LT)
and the rescaled unitarity triangle (UT), which
provides an intuitive and scale-independent connection of
$CP$-violating observables to quark mass ratios, is
particularly worth mentioning. 

Let us make some further comments on the quark mass matrix
(4.3), its phenomenological hints and its theoretical
prospects. 

Naively one might not expect any prediction from the 
four-texture-zero mass matrices in (4.3), 
since they totally consist of ten free parameters
(two of them are the phase differences between $M_{\rm u}$ and $M_{\rm d}$).
This is not true, however, as we have seen. 
We find that two predictions,
$\tan\theta_{\rm u} \approx \sqrt{m_u/m_c}$ and $\tan\theta_{\rm d} \approx 
\sqrt{m_d/m_s}$ , can be obtained in the leading order approximation. 
In some cases
the latter may even hold in the next-to-leading order 
approximation, as shown in (4.4) and (4.9). 
Note again that these two relations,
as a consequence of the hierarchy and texture zeros of our
quark mass matrices, are essentially independent of the renormalization-group 
effects. This interesting scale-independent
feature can also be seen from the LT and the rescaled
UT as well as their
inner angles $(\alpha, \beta, \gamma)$. 

It remains to be seen whether the interesting possibility
$\varphi \approx \phi_1 \approx 90^{\circ}$, indicated by current
data of quark masses and flavor mixing, could arise from an
underlying flavor symmetry or a dynamical
symmetry breaking scheme. Some speculations about this problem 
have been made 
(see, e.g., Refs. \cite{FX95} and Refs. \cite{Democracy,New}).
However, no final conclusion has been reached thus far.
It is remarkable, nevertheless, that we have at least observed a
useful relation between the area of the UT
(${\cal A}_{\rm UT}$) and that of the LT
(${\cal A}_{\rm LT}$) to a good degree of accuracy:
\begin{equation}
{\cal A}_{\rm UT} \; \approx \; |V_{cb}|^2 {\cal A}_{\rm LT}
\; \approx \; \sin^2\theta ~ {\cal A}_{\rm LT} \; .
\end{equation}
Since ${\cal A}_{\rm UT} = {\cal J}/2$ measures the magnitude
of $CP$ violation in the standard model, we conclude that 
$CP$ violation is primarily linked to the 
light quark sector. This is a natural consequence of the
strong hierarchy between the heavy and light quark masses,
which is on the other hand responsible for the smallness of 
${\cal J}$ or ${\cal A}_{\rm UT}$. 

Is it possible to derive the quark mass matrix (4.3) in some
theoretical frameworks? To answer this question
we first specify the hierarchical structure of $M_{\rm q}$ in 
terms of the mixing angle $\theta_{\rm q}$ (for q = d or s). 
Adopting the radiant
unit for the mixing angles (i.e., $\theta_{\rm u}
\approx 0.085$, $\theta_{\rm d} \approx 0.204$ and
$\theta \approx 0.040$), we have
\begin{eqnarray}
\frac{m_u}{m_c} & \sim & \frac{m_c}{m_t} \;
\sim \; \theta^2_{\rm u} \; ,  \nonumber \\
\frac{m_d}{m_s} & \sim & \frac{m_s}{m_b} \;
\sim \; \theta^2_{\rm d} \; .
\end{eqnarray}
Then the mass matrices $M_{\rm u}$ and $M_{\rm d}$,
which have the mass  scales $m_t$  and $m_b$ 
respectively, take the following {\it parallel} 
hierarchies:
\begin{eqnarray}
M_{\rm u} & \sim & m_t \left ( \matrix{
{\bf 0}	& \theta^3_{\rm u}	& {\bf 0} \cr
\theta^3_{\rm u} & \theta^2_{\rm u} & \theta^2_{\rm u} \cr
{\bf 0}	& \theta^2_{\rm u}	& {\bf 1} \cr}
\right ) \; \; , \nonumber \\
\nonumber \\
M_{\rm d} & \sim & m_b \left ( \matrix{
{\bf 0} & \theta^3_{\rm d}	& {\bf 0} \cr
\theta^3_{\rm d} & \theta^2_{\rm d} & \theta^2_{\rm d} \cr
{\bf 0} & \theta^2_{\rm d}	& {\bf 1} \cr}
\right ) \; \; ,
\end{eqnarray}
where the relevant complex phases have been neglected.
Clearly all three flavor mixing angles can properly
be reproduced from (5.3), once one takes $\theta \approx
\theta^2_{\rm d} \gg \theta^2_{\rm u}$ into account. 
The $CP$-violating phase $\varphi$ in $V$ comes
essentially from the phase difference between the
$\theta^3_{\rm u}$ and $\theta^3_{\rm d}$ terms.

Of course $\theta_{\rm u}$ and $\theta_{\rm d}$, which
are more fundamental than the Cabibbo angle $\theta_{\rm C}$
in our point of view, denote perturbative corrections to the
rank-one limits of $M_{\rm u}$ and $M_{\rm d}$
respectively. They are responsible for the generation 
of light quark masses as well as the flavor mixing.
They might also be responsible for $CP$ violation in
a specific theoretical framework (e.g., the pure
real $\theta_{\rm u}$ and the pure imaginary $\theta_{\rm d}$
might lead to a phase difference of about $90^{\circ}$
between $M_{\rm u}$ and $M_{\rm d}$, which is just the
source of $CP$ violation favored by current data).
The small parameter $\theta_{\rm q}$ could get its
physical meaning in the Yukawa coupling of
an underlying superstring theory: 
$\theta_{\rm q} = \langle \Theta_{\rm q} \rangle /\Omega_{\rm q}$,
where $\langle \Theta_{\rm q} \rangle $ denotes the
vacuum expectation value of the singlet field $\Theta_{\rm q}$,
and $\Omega_{\rm q}$ represents  the unification (or string)
mass scale which governs higher  dimension operators
(see, e.g., Ref. \cite{Froggatt79}). The quark mass
matrices of the form (5.3) could then be obtained 
by introducing an extra (horizontal) U(1) gauge symmetry
or assigning the matter fields appropriately.

A detailed study of possible dynamical models responsible 
for the quark mass matrices (4.3) or (5.3) is certainly desirable.
We believe that
the texture zeros and parallel hierarchies of up and down
quark mass matrices do imply specific symmetries, perhaps 
at a superhigh scale, and have instructive consequences
on flavor mixing and $CP$-violating phenomena.
The new parametrization of the flavor mixing matrix
that we advocated is particularly useful in studying
the quark mass generation, flavor mixing 
and $CP$ violation. 


\end{document}